\documentstyle[sprocl]{article}
\input{psfig}
\def\be{\begin{equation}}
\def\ee{\end{equation}}
\def\bea{\begin{eqnarray}}
\def\eea{\end{eqnarray}}
\def\simlt{\stackrel{<}{{}_\sim}}
\def\simgt{\stackrel{>}{{}_\sim}}
\newcommand{\ovm}{{\overline m}}
\begin{document}
\begin{flushright}
CERN-TH/97-28\\
hep--ph/9702431 \\
\end{flushright}
\title{WHAT IS THE MASS OF THE LIGHTEST SUPERSYMMETRIC HIGGS BOSON?
\footnote{To appear in ``Perspectives on Higgs Physics II'' 
ed. G.L. Kane, World Scientific, Singapore.}}
\author{PIOTR H. CHANKOWSKI \footnote{On leave of absence from the
Institute of Theoretical Physics, Warsaw University, Ho\.za 69, 
00-681 Warsaw, Poland.}}
\address{Theory Division, CERN, Geneva 23, Switzerland}
\author{STEFAN POKORSKI}
\address{Inst. of Theoretical Physics, Warsaw University, Ho\.za 69,
00-681 Warsaw, Poland}
\maketitle\abstracts{After reviewing briefly the upper bounds on the mass
of the lightest Higgs boson in the most general unconstrained
minimal supersymmetric extension of the Standard Model, we discuss
various arguments which reduce the parameter space of the model and 
give stronger predictions for $M_{h^0}$. First,  the constraints from the
presently available experimental data are summarized. Next, the role of 
of several additional theoretical assumptions is studied, after
extrapolating the model to high energy scales. The most important ones 
are: perturbative validity up to the GUT scale and the electroweak
symmetry breaking. 
A Higgs boson with $M_{h^0}<$100 GeV  is predicted in several scenarios.
Its absence in that mass range will
have important implications for the parameter space of the model.}

\section{Introduction}
Spontaneous breaking of the electroweak gauge symmetry $SU(2)\times U(1)$
is now confirmed experimentally with one per mille accuracy (see the Chapter
by A. Blondel). However, the actual mechanism of this symmetry breaking
still remains unknown and waits for experimental discovery. This is, by far,
the most central question to particle physics and, in particular, to the 
experimental programs at LEP 2 and the LHC. 
It is very likely that the understanding of
the mechanism of the electroweak symmetry breaking will not only provide
us with the missing link in the Standard Model but, also, will be an 
important bridge to physics beyond it.

The minimal model for spontaneous electroweak gauge symmetry breaking is the
Higgs mechanism, whose minimal version (the minimal Standard Model) requires
one scalar $SU(2)$ doublet (Higgs doublet). The scalar potential at some
scale $\Lambda$ is:
\begin{eqnarray}
V(\Lambda) = m^2(\Lambda) |H(\Lambda)|^2 + 
{1\over2}\lambda(\Lambda)|H(\Lambda)|^4
\end{eqnarray}
with the dependence on $\Lambda$ controlled by the renormalization group
evolution (RGE). The mass of the physical scalar (Higgs boson) is
$M^2_{\phi^0}= \lambda(M_Z)v^2(M_Z)$ where 
$v(M_Z)=\sqrt{4M^2_W/g^2_2(M_Z)}\approx246$ GeV. There exist the well known 
theoretical bounds on the Higgs boson mass (see \cite{QUIROS} for an extensive
discussion) which follow from certain constraints on the behaviour of the 
self-coupling $\lambda(\Lambda)$. One can distinguish two types of bounds. The 
most general upper bound on $M_{\phi^0}$ follows from the requirement that the
Standard Model is a unitary and weakly interacting theory at the energy scale 
${\cal O}(M_Z)$. We get then $M_{\phi^0}\simlt{\cal O}(1$ TeV). Stronger 
bounds are $\Lambda$-dependent and are known under the names of the triviality 
(upper) bound and the vacuum stability (lower) bound. They follow respectively
from the requirements that the theory remains perturbative 
($\lambda(\Lambda)<16\pi^2$) and the vacuum remains stable
($\lambda(\Lambda)>0$) up to a certain scale $\Lambda$. Those bounds are 
particularly interesting in the presence of the heavy top quark, 
$m_t=(175\pm6)$ GeV. They are shown in Fig. 3 in ref. \cite{QUIROS} and lead 
to the striking 
conclusions: We see that the discovery of a light Higgs boson 
($M_{\phi^0}\simlt80$ GeV) or a heavy one ($M_{\phi^0}\simgt500$ GeV) would be 
a direct information about the existence of new physics below the scale
$\Lambda\sim {\cal O}(1$ TeV) (or at least of a strongly interacting Higgs 
sector). On the other hand, if the SM in its perturbative regime is to be 
valid up to very large scales $\Lambda$, of the order of the GUT scale 
$\Lambda\approx10^{16}$ GeV, one gets strong bounds 140 GeV 
$\simlt M_{\phi^0}\simlt180$ GeV. In this case we face the well known hierarchy
problem in the SM: $M_{\phi^0}$, $v$ $\ll\Lambda$ and it is difficult to 
understand how the scalar potential remains stable under radiative corrections 
of the full theory. One way or another, the bounds on $M_{\phi^0}$ in the SM 
are strongly suggestive that the mechanism of spontaneous electroweak gauge 
symmetry breaking is directly related to the existence of a new scale (not 
much above the electroweak scale) in fundamental interactions. The central 
question can then be phrased as this: discover and investigate the next scale 
in fundamental interactions. Is it the scale of new strong interactions 
(strongly interacting Higgs sector or techicolour interactions or compositness
scale) or the scale of soft supersymmetry breaking? We would like to stress the
basic difference between these two lines of approach. In the first one, the 
new scale is also a cut-off scale for the perturbative validity of the 
electroweak theory. Supersymmetry offers a solution to the hierarchy problem 
while maintaining the perturbative nature of the theory up to the GUT or even 
Planck scale. This is a welcome feature if such facts as the gauge coupling 
unification are not to be considered as purely accidental.

Another difference is in the expectations for the Higgs boson mass: in the 
strong interaction scenarios it is naturally heavy, with its mass close to 
the new scale $\Lambda$. In supersymmetric extensions of the SM the lightest 
Higgs scalar $h^0$ generically remains light, $M_{h^0}\sim {\cal O}(100$ GeV) 
and only logarithmically correlated with the scale of the soft supersymmetry 
breaking. 

It is the purpose of this Chapter to summarize the predictions for the Higgs 
sector in supersymmetric extensions of the SM. The Higgs sector in the 
Minimal Supersymmetric Standard Model (MSSM), considered as a low energy 
effective theory with all free parameters totally unconstrained has been 
discused in detail in \cite{HABER}. For the sake of easy reference and to 
establish our notation we summarize those results in Section 2. The main 
conclusion of that Section is the specification of the set of parameters which
determine the Higgs boson masses and the existence of general upper 
bounds for the mass of the lightest supersymmetric Higgs boson $h^0$. Next, 
in the main part of this article, we collect the available results and 
arguments which constrain the general parameter space relevant for the Higgs 
mass in the MSSM and, therefore, give more specific predictions for $M_{h^0}$. 
One should stress that our interest in the MSSM is well motivated.
It is structured in such a way that the success of the SM in describing the 
precision electroweak data is maintained. Moreover, its virtue is that it can 
be extrapolated up to the large energy scales (the scales where the
soft supersymmetry breaking terms are generated) in unambigous  and 
quantitative way. We shall mainly consider the supergravity scenario for
supersymmetry breaking in which the MSSM is extrapolated up to the GUT scale.
A brief discussion of the gauge mediated symmetry breaking models 
\cite{DIFISR} is, however, also included. Broadly speaking, the weak scale - 
large scale connection is the main source of constraints on the Higgs sector 
parameter space which we are going to present. Finally, in Section 5,
we discuss the predictions for the Higgs sector in non-minimal
versions of the supersymmetric extensions of the Standard Model. 

\section{Higgs sector in MSSM - a brief summary}
In the Minimal Supersymmetric Standard Model the Higgs sector is particularly 
simple and predictive. Supersymmetry and the minimal particle content imply
that it consists of two Higgs doublets, each coupled to only one type 
($H_1(H_2)$ couples to the down (up)) of fermions. 
The scalar Higgs potential reads:
\begin{eqnarray}
V = m^2_1 \overline H_1H_1 + m^2_2 \overline H_2H_2
+ m_3^2 \left(\epsilon_{ab}H_1^aH_2^b + c.c\right)\nonumber\\
+{1\over8}(g_1^2+g_2^2)(\overline H_1H_1-\overline H_2H_2)^2
+{1\over2}g^2_2|\overline H_1H_2|^2
\label{eqn:treepot}
\end{eqnarray}
where $\epsilon_{12}=-1$ and $m_1^2$, $m_2^2$ and $m_3^2$ are the soft 
supersymmetry breaking mass parameters. The crucial point about the potential 
(\ref{eqn:treepot}) is that its quartic couplings are the electroweak gauge 
couplings (i.e. there is no $F$-term contribution to the scalar Higgs 
potential). The only free parameters are the three mass parameters. The tree 
level mass eigenstates of the Higgs bosons are: two $CP$-even ($h^0$, $H^0$), 
one $CP$-odd ($A^0$) and 2 charged ($H^\pm$) physical particles and three 
Goldstone bosons ``eaten up'' by the gauge bosons. An important parameter is
$\tan\beta\equiv v_2/v_1$ where $v_i$ minimize the tree level potential 
(\ref{eqn:treepot}) and are given by $v_1=v\cos\beta$, $v_2=v\sin\beta$ with
\begin{eqnarray}
v^2={8\over g^2_1+g_2^2}{m^2_1 - m^2_2\tan^2\beta\over\tan^2\beta-1}
\label{eqn:vsquared}
\end{eqnarray}
\begin{eqnarray}
\sin2\beta={2m^2_3\over m^2_1+m^2_2}
\label{eqn:sinbeta}
\end{eqnarray}
Since $v$ is fixed by the $Z^0$ mass, all physical Higgs boson masses are 
expressed in terms of only two free parameters. They can be taken e.g. as 
$\tan\beta$ and the mass $M^2_{A^0}$ of the $CP-$odd Higgs scalar $A^0$ 
given by $M^2_{A^0}=m^2_1+m^2_2$. The $CP-$even Higgs boson masses then read:
\begin{eqnarray}
M^2_{h^0,H^0}={1\over2}\left(M^2_{A^0} + M^2_{Z^0} +
\sqrt{\left(M^2_{A^0}-M^2_{Z^0}\right)^2-
4M^2_{A^0}M^2_{Z^0}\cos^22\beta} \right)
\label{eqn:treemass}
\end{eqnarray}
leading to the bound $M_{h^0}<M_{Z^0}$ and to the ``natural'' (i.e. independent
of any other parameters) relation $M^2_{h^0}+M^2_{H^0}=M^2_{A^0}+M^2_{Z^0}$.
The other relation is $M^2_{H^\pm}=M^2_{W^\pm} + M^2_{A^0}$ \cite{LISH}.

The origin and the magnitude of radiative corrections to the Higgs boson
masses can be easily understood. Let $M$ be the scale of the soft supersymmetry
breaking sfermion masses. Neglecting terms suppressed by inverse powers of 
$M$, the dominant one-loop corrections to the effective potential $V_{eff}$,
due to the top and stop loops, can be absorbed into renormalization of the 
parameters in the Higgs potential. One gets:
\begin{eqnarray}
V = \tilde m^2_1 \overline H_1H_1 + \tilde m^2_2 \overline H_2H_2
+ \tilde m_3^2 \left(\epsilon_{ab}H_1^aH_2^b + c.c\right)\nonumber\\
+ \lambda_1 |H_1|^4 +\lambda_2 |H_2|^4 +\lambda_3 |H_1|^2 |H_2|^2
+\lambda_4 |\overline H_1H_2|^2
\label{eqn:corrpot}
\end{eqnarray}
The appearence of other couplings is protected by the symmetries of the model.
It is clear on the dimensional grounds that 
\begin{eqnarray}
\delta m^2_i = \tilde m^2_i - m^2_i \sim {\cal O}(M^2)
\end{eqnarray}
They are logarithmically divergent but can be absorbed into the free parameters
of the model. The corrections $\delta\lambda_i$ defined by
\begin{eqnarray}
\delta\lambda_1 = \lambda_1 - {1\over8}(g_1^2+g_2^2), ~~~~~~~~
\delta\lambda_2 = \lambda_2 - {1\over8}(g_1^2+g_2^2)\nonumber\\
\delta\lambda_3 = \lambda_3 + {1\over4}(g_1^2+g_2^2), ~~~~~~~~
\delta\lambda_4 = \lambda_4 - {1\over2}g_2^2
\end{eqnarray}
are all ${\cal O}(\log M)$. Moreover, from the non-renormalization theorem,
the corrections $\delta\lambda_i$ are calculable (finite) in terms of the
remaining parameters of the model. From the top and stop loops with attached 
four Higgs boson legs one gets
\begin{eqnarray}
\delta\lambda_i\sim {12\over16\pi^2}
h^4_t\log\left({M^2_{\tilde t}\over m_t^2}\right)
\end{eqnarray}
where $h_t$ is the top quark Yukawa coupling (factor 12 comes from 4 top 
degrees of freedom multipied by the color factor of 3) and $M_{\tilde t}$
denotes the scale of the stop masses. Thus, the correction to the $h^0$ 
mass is \cite{OKYAYA}
\begin{eqnarray}
\delta M^2_{h^0}\sim 
{\cal O}\left({6g_2^2\over16\pi^2}{m_t^4\over M^2_W}
\log\left({M^2_{\tilde t}\over m_t^2}\right)\right)
\end{eqnarray}
In general, taking into account the full structure of the stop mass 
matrix, the lightest Higgs boson mass in the MSSM is parametrized by 
\begin{eqnarray}
M_{h^0} = M_{h^0}
\left(M_{A^0},\tan\beta,m_t,M_{\tilde t_1},M_{\tilde t_2},A_t,\mu,...\right)
\end{eqnarray}
where $M_{\tilde t_i}$ are the physical stop masses, $A_t$ and $\mu$ determine
their mixing angle (as well as some of their trilinear couplings to the Higgs
bosons) and ellipsis stand for other parameters whose effects are not dominant
(e.g. the sbottom sector parameters). 

\begin{figure}
\psfig{figure=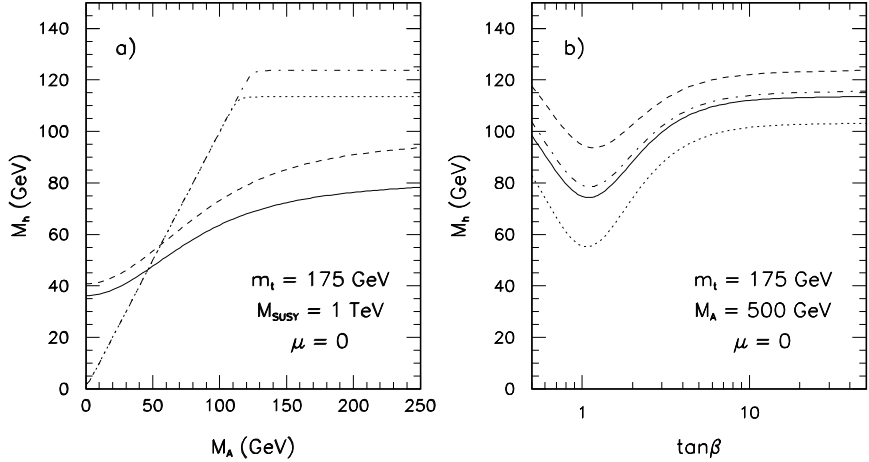,width=12.0cm,height=7.cm}
\caption{Radiatively corrected $M_h$ in the MSSM (1- and 2-loops). {\bf a)} 
As a function of the $CP-$odd Higgs mass for $M_{SUSY}=1$ TeV and for 
$\tan\beta=1.5$ (solid and dashed lines) and  $\tan\beta=50$ (dotted and 
dash-dotted lines). Lower (upper) lines correspond to $A_t=0$ 
$(2.5 M_{SUSY})$ ~{\bf b)} As a function of $\tan\beta$ for $m_Q=m_U=1$ TeV, 
$A_t=0$ (2.5 TeV) solid (dashed) line and for $m_Q=500$, $m_U=100$ GeV, 
$A_t=0$ (1 TeV) dotted (dash-dotted) line.
\label{fig:mh}}
\end{figure}

In Fig. \ref{fig:mh}a we show
$M_{h^0}$ as a function of $M_{A^0}$ for two values of $\tan\beta$ 
and $M_{\tilde t_1}=M_{\tilde t_2}=1$ TeV, $\mu=0$ and two values of the
$A_t$ parameter. We see that
maximal $M_{h^0}$ is always obtained for $M_{A^0}\gg M_{Z^0}$ (in practice, 
the bound is saturated for  $M_{A^0}\simgt250$ GeV).
In this limit one gets from the effective potential
approach (see ref. \cite{HABER} for details) particularly simple
result for the one-loop corrected $M_{h^0}$ \cite{HE}:
\begin{eqnarray}
M^2_{h^0}=M^2_{Z^0}\cos^22\beta+{3\alpha\over4\pi s^2_W}{m^4_t\over M^2_W}
\left[\log\left({M^2_{\tilde t_1}M^2_{\tilde t_2}\over m^4_t}\right)
+\left({M^2_{\tilde t_1}-M^2_{\tilde t_2}\over2m^2_t}\sin^22\theta_{\tilde t}
\right)^2\right.\nonumber
\end{eqnarray}
\begin{eqnarray}
\times\left. f(M^2_{\tilde t_1},M^2_{\tilde t_2})
+ {M^2_{\tilde t_1}-M^2_{\tilde t_2}\over2m^2_t}\sin^22\theta_{\tilde t}
\log\left({M^2_{\tilde t_1}\over M^2_{\tilde t_2}}\right)\right]
\label{eqn:mhha}
\end{eqnarray}
where $f(x,y)=2 - (x + y)/(x - y) \log(x/y)$. For large $M_{A^0}$ the separate 
dependence on the parameters $A_t$ and $\mu$ has disappeared and is replaced 
by the effective dependence on the left-right stop mixing angle 
$\theta_{\tilde t}$. One should also mention (see \cite{HABER}) that the 
two-loop corrections to $M_{h^0}$ are typically ${\cal O}(20$\%) of 
the one-loop corrections and are negative. In Figs. \ref{fig:mh} and 
\ref{fig:bsgamma} they are taken into account in the approach proposed in 
\cite{CAESQUWA}.The dependence on $\tan\beta$, illustrated in Fig. 
\ref{fig:mh}b, is important for our further discussion. 

\section{Experimental constraints on the parameters of the MSSM}
Having recalled the general parameter set relevant for the Higgs sector in 
supersymmetric models we proceed now to discuss constraints on 
those parameters which follow from various additional considerations. 

The parameters of the MSSM which enter into the prediction for $M_{h^0}$
are partially constrained by other experimental data. Although at present
relatively weak, those are interesting constraints which correlate in the
framework of the MSSM the Higgs boson mass(es) with other measurements
without any additional assumptions. There are also stronger constraints 
which follow from embedding the MSSM into a SUSY GUT scenario, or more
generally, from extrapolating the MSSM up to the GUT scale, supplemented 
with several plausible (simple) assumptions about physics at that scale.
We begin our discussion with the former.

\begin{figure}
\psfig{figure=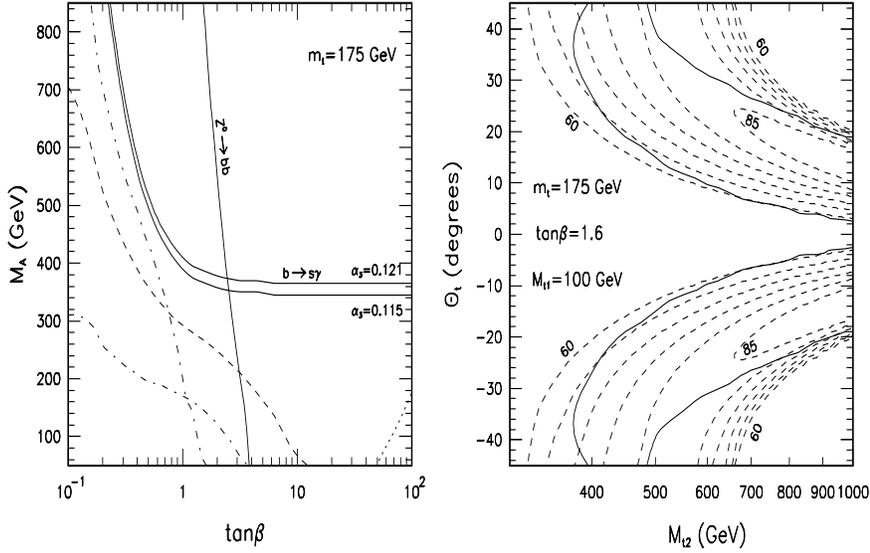,width=12.0cm,height=8.0cm}
\caption{{\bf a)} Bounds on the $CP$-odd Higgs boson mass in the MSSM 
with very heavy spectrum for $\alpha_3=$0.115 and 0.121 (solid lines)
and (for $\alpha_3=$0.117) for the chargino and stop masses 500 and 250 GeV -
the dashed and dash-dotted lines respectively. The dotted line shows the 
constraint from $b\rightarrow c\tau\overline\nu_{\tau}$. Also shown 
are the constraints from $Z^0\rightarrow\overline bb$ in the 
MSSM with heavy spectrum (solid line) and for the chargino and stop masses 
equal 250 GeV (dashed line).
{\bf b)} Contours of the constant Higgs boson mass 
$M_{h^0}=60$, 65, 70, 75, 80 and 85 in the plane 
($M_{\tilde t_2}$, $\theta_{\tilde t}$) for $M_{A^0}=1000$ GeV, 
$\tan\beta=1.6$ and the lighter stop mass equal 100 GeV.
The region where $\Delta\chi^2<4$ 
(see the text) is bounded by the solid lines.
\label{fig:bsgamma}}
\end{figure}

A number of experimental data constrains the tree level parameters $M_{A^0}$
and $\tan\beta$. These are, first of all, $BR(b\rightarrow s\gamma)$ 
\cite{BUMIMUPO}, $\Gamma(Z^0\rightarrow\overline bb)$ \cite{BUFI}
and (for large $\tan\beta$) $BR(b\rightarrow c\tau\overline\nu_{\tau})$
\cite{KRPO}. Other processes (e.g. 
$B^0$-$\overline B^0$ mixing) give weaker constraints. The first two processes
are sensitive to the $M_{H^\pm}$ and $\tan\beta$ via the charged Higgs boson 
- top quark loop contribution and the third one via the tree level $H^{\pm}$ 
exchange. The $btH^-$ coupling is given by
\begin{eqnarray}
{\cal L}_{btH^\pm} ={g_2\over\sqrt2M_W}H^+ 
\overline\psi_t \left(m_t\cot\beta P_L + m_b\tan\beta P_R\right)V_{tb}\psi_b
+ h.c.
\label{eqn:coupl1}
\end{eqnarray}
where $P_{L,R}\equiv(1\mp\gamma^5)/2$ (the coupling $bcH^\pm$ is similar
with $\psi_t\rightarrow\psi_c$ and $V_{tb}\rightarrow V_{cb}$) and the
coupling $\tau\nu_\tau H^\pm$ is given by:
\begin{eqnarray}
{\cal L}_{\tau\nu_\tau H^\pm} ={g_2\over\sqrt2}{m_\tau\over M_W}\tan\beta H^+ 
\overline\psi_{\nu_\tau} P_R\psi_\tau
+ h.c.
\label{eqn:coupl}
\end{eqnarray}

\noindent We see that  $BR(b\rightarrow c\tau\overline\nu_{\tau})$ can 
be enhanced for large $\tan\beta$ and light $H^\pm$ and the measurement 
$BR(b\rightarrow c\tau\overline\nu_{\tau})=2.69\pm0.44$\% gives the bound
\cite{KRPO} 
\begin{eqnarray}
\tan\beta\simlt0.52{M_{H^\pm}\over 1 ~{\rm GeV}}
\end{eqnarray}
which is essentially independent of the other parameters of the MSSM
and is shown in Fig. \ref{fig:bsgamma}a by the dotted line. 

The first two processes get contributions also from diagrams with 
superpartners in the loop (the dominant one may come from the chargino-stop
loop, due to the large Yukawa coupling for the higgsino component with the 
right-handed stop) and, in consequence, the exclusion regions in the 
$(M_{A^0}, \tan\beta)$ plane depend on supersymmetric parameters. The 
strongest dependence is  on the $m_{C^\pm}$, $M_{\tilde t_i}$, $\theta_t$.
In the limit of very heavy superpartners ($m_{C^\pm_i}$, 
$M_{\tilde t_i}\simgt$1 TeV) we get the exclusion 
limits shown in Fig. \ref{fig:bsgamma}a by the solid curves. 
The dependence of these limits on $m_{C^\pm_1}$, $M_{\tilde t_1}$ 
(with all other superpartners at 1 TeV) is illustrated by the dashed 
(dot-dashed) curve which is obtained for 
$m_{C^\pm_1}=M_{\tilde t_1}=$500 (250) 
GeV. In each case a scan over the chargino composition and the left-right
mixing angle in the stop sector have been performed.

Negative contribution of a light charged Higgs boson - top quark loop to
$R_b\equiv\Gamma(Z^0\rightarrow\overline bb)/\Gamma(Z^0\rightarrow hadrons)$
becomes too large for small values of $\tan\beta$ (see eq. (\ref{eqn:coupl1}))
\cite{BUFI} excluding, in the MSSM with heavy spectrum, the region of the 
($\tan\beta$, $M_{A^0}$) plane to the left of the almost vertical solid line 
shown in Fig. \ref{fig:bsgamma}a (we require that $R_b$ remains 
within $2\sigma$ of the presently measured value $R_b^{exp}=0.2179\pm0.0012$ 
\cite{LEPEWWG}). Positive contribution of the chargino -  stop loop 
weakens significantly this bound (if both are light) as shown for 
$m_{C_1^\pm}=M_{\tilde t_1}=$250 GeV by the almost vertical dot-dashed line.
Similar effects are observed for $BR(b\rightarrow s\gamma)$. In the limit 
of heavy superpartners one obtains strong bound
\footnote{We require the branching ratio computed in the NLO 
approximation to fall into the interval 
$1\times10^{-4}<BR(b\rightarrow s\gamma)<4\times10^{-4}$. 
All uncertainties of the computation (for detailed discussion of the
uncertainties see ref. \cite{CHMIMU}) are taken into account.}
on $M_{A^0}$: the $H^\pm$ - 
top loop adds positively to the SM contribution which, by itself, is in the
upper edge of the experimentally allowed region. The bound is weakened 
in the presence of a light chargino - stop loop (which can interfere negatively
with the $H^\pm$ -top loop) and even totally disappears for large values
of $\tan\beta$. One should note, however, that for $\tan\beta\sim m_t/m_b$
the interference term is generically very large and consistency with the
data requires a large amount of fine-tuning in the 
$(\theta_{\tilde t}, M_{\tilde t_1})$ parameter space \cite{CHPO}
\footnote{Those results are obtained under the assumption that the chargino
-stop - bottom coupling is given by the Kobayashi-Maskawa angles. 
In principle, off-diagonal terms in the right-handed stop mass matrix 
are possible which would change those couplings. Such (1,3) and (2,3) 
terms are not constrained by any other data. However, other flavour off 
diagonal terms are strongly constrained so the presence of large (1,3) and 
(2,3) terms for right-handed squarks would mean strong flavour dependence 
in the squark mass matrices.}.
Thus, one concludes that the large $\tan\beta$ scenario is unlikely to be
consistent with a light $CP$-odd Higgs boson.  For $M_{h^0}$ this implies
the plateau region in Fig. \ref{fig:mh}a. (The present experimental 
bound \cite{ALEPH} is $M_{A^0}\simgt55$ GeV for $\tan\beta\simgt50$.)

Radiative corrections to $M_{h^0}$ 
are mainly dependent on the stop masses (for large
$\tan\beta$ also on the sbottom mass) and on the parameters $A_t$ and $\mu$
(in the large $M_{A^0}$ limit they depend only on the combination
$A_t+\mu\cot\beta$ which can be traded for $\theta_{\tilde t}$ as in eq.
(\ref{eqn:mhha})) which are constrained by the precision data. A light
left-handed stop would introduce additional source of the custodial $SU_V(2)$
breaking. Since the SM fit to the LEP precision data is very good with the
$SU_V(2)$ breaking given mainly by the $t$-$b$ mass spliting, additional 
sources of the custodial $SU_V(2)$ breaking would tend to destroy the quality
of the SM fit. In Fig. \ref{fig:bsgamma}b we show by the solid lines
the limits in the ($M_{\tilde t_2}$, $\theta_{\tilde t}$) plane for 
$M_{\tilde t_1}=100$ GeV 
obtained from the requirement that $\Delta\chi^2<4$ compared with the minimum
of a fit to the electroweak observables  found for heavy $\tilde t_2$.
Similar bounds exist for heavier $\tilde t_1$.
In Fig. \ref{fig:bsgamma}b, those limits are shown together with the 
contours of constant $M_{h^0}$.

In summary, the present experimental data 
do not significantly improve the general upper bound on
$M_{h^0}$. However they give constraints on SUSY parameters and 
there are interesting correlations between parameter regions allowed by other 
processes (e.g. a light $CP$-odd Higgs boson is consistent with 
$BR(b\rightarrow s\gamma)$ and $\Gamma(Z^0\rightarrow\overline bb)$ 
only if chargino and stop are also light) and the Higgs boson mass $M_{h^0}$.

\section{The weak scale - large scale connection}
The parameter space of the low energy MSSM can be further reduced by 
introducing additional theoretical ideas. The first one involves the 
extrapolation of the MSSM up to very high energy scales (the ``desert''
hypothesis) and the observation that to a very good approximation the 
$SU(3)\times SU(2)\times U(1)$ gauge couplings converge to a common value.
With the supersymmetry breaking scale of order 1 TeV or less the unification 
takes place at an energy scale of order $10^{16}$ GeV and depends weakly on
the details of the GUT-scale theory. One can rephrase this result by saying
that e.g. the Weinberg angle is correctly predicted in terms of the measured
values of $\alpha$ and $\alpha_3$ from the hypothesis of the gauge coupling
unification. This is one of the most compelling hints for the low energy 
supersymmetry and, this should be strongly stressed, for the fact that 
physics remains perturbative up to the GUT scale $\sim10^{16}$ GeV. Once
assumed, the perturbative validity 
of the MSSM up to the scale $\sim10^{16}$ GeV
has several interesting implications for the behaviour of the third
generation Yukawa couplings and for the ``interesting'' (i.e. most plausible?)
values of $\tan\beta$ and, in consequence, also for $M_{h^0}$. 

The first important notion is that of the quasi-infrared fixed point for the
top quark (or top and bottom quarks for large $\tan\beta$) Yukawa coupling. 
We recall the fixed point structure of the top quark Yukawa coupling $Y_t$ ~
($Y_t\equiv h_t^2/4\pi$) in the MSSM. The renormalization group equations
have the form:
\begin{eqnarray}
{dY_t\over dt}&=& Y_t (a_i^u\alpha_i - c_tY_t)\nonumber\\
{d\alpha_i\over dt}&=&-b_i\alpha_i^2
\label{eqn:yukrge}
\end{eqnarray}
where $2\pi t=\log (M_{GUT}/Q)$, ~
$a_i^u=(13/15,3,16/3)$, $b_i=(11,1,-3)$, $c_t=6$  and
$\alpha_i=g_i^2/4\pi$. Ignoring the smaller electroweak couplings, $Y_t$ is 
related to the QCD coupling $\alpha_3$: the fixed point solution for the 
ratio $Y_t/\alpha_3$ reads \cite{PERO}:
\begin{equation}
Y_t^F(t) = {a^u_3 + b_3\over c_t}\alpha_3(t)
\end{equation}

One can also solve eqs. (\ref{eqn:yukrge}) explicitly \cite{IBLO}:
\begin{equation}
Y_t(t) = {4\pi Y_t(0)E(t)\over4\pi+c_tY_t(0)F(t)}
\end{equation}
with 
\begin{equation}
E(t)\approx\left({\alpha_3(0)\over\alpha_3(t)}\right)^{{a^u_3\over b_3}},~~~~~
F(t) = \int_0^t E(t')dt'
\end{equation}
It may happen that $Y_t^F(t)$ is not reached because of too short
a running but, nevertheless, $c_tY_t(0)F(t)\gg4\pi$ and
\begin{equation}
Y_t(t)\approx Y_t^{QF}(t) = {4\pi E(t)\over c_tF(t)}
\end{equation}
i.e.the low energy value of $Y_t$ no longer depends on the initial
value $Y_t(0)$. This is called the quasi-infrared fixed point solution (Q-IR)
\cite{HI} and we have:
\begin{equation}
Y_t^{QF}(t)\approx  Y_t^F(t){1\over1-\left({\alpha_3(t)\over\alpha_3(0)}
\right)^{1+ {a^u_3\over b_3}}}
\end{equation}

\begin{figure}
\psfig{figure=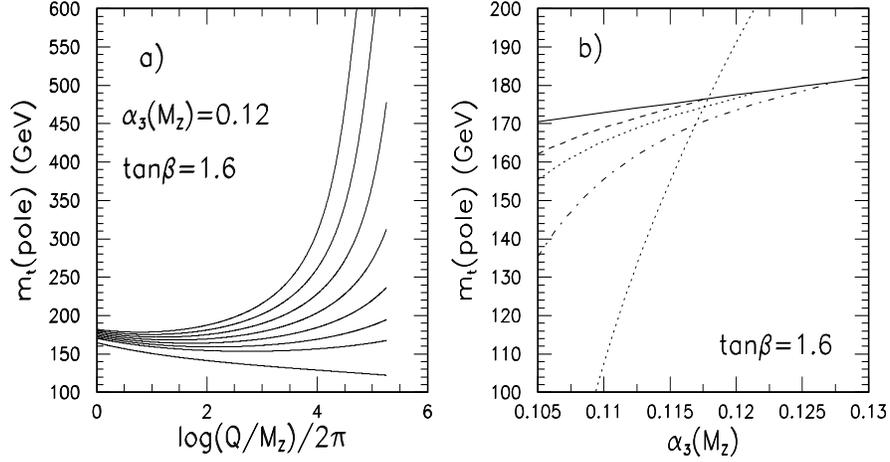,width=12.0cm,height=7.5cm}
\caption{{\bf a)} Top quark mass RG running for 
different boundary conditions at the scale $2\times10^{16}$ GeV. ~${\bf b)}$ 
Q-IR top quark mass
prediction (solid line) and top quark masses necessary to achieve 
bottom-tau Yukawa coupling unification for $m_b(pole)=$4.7 (dashed line),
4.9 (dotted) and 5.2 (dot-dashed) GeV. 
Gauge coupling unification favours the region to the right of the second 
dotted line.
\label{fig:irfix}}
\end{figure}

This situation indeed occurs in the MSSM (for $b_3$, $a^u_3$, $c_t$ of the 
MSSM) for $Y_t(0)\simgt\cal O$(0.1) (see Fig. \ref{fig:irfix}a)
i.e. for the initial  values still in the
perturbative regime! Thus, not only $Y_t^{QF}(M_Z)$ is the upper bound
for $Y_t(M_Z)$ but it can be  reached at the limit of perturbative
physics \cite{ALPOWI}. The Q-IR prediction
for the running top quark mass in the ${\overline MS}$ 
scheme, for $\alpha_3(M_Z)$=
0.11-0.13 and small and moderate values of $\tan\beta$ 
is approximately given by \cite{CAWA}:
\begin{equation}
m_t^{QF}(m_t)\approx196 {\rm GeV}\left[1+2(\alpha_3(M_Z)-0.12)\right]\sin\beta
\label{eqn:mtopir}
\end{equation}

\noindent
The physical top quark mass (pole mass) is obtained by including QCD
corrections which contribute ${\cal O}$(10 GeV) to the final result.
Eq. (\ref{eqn:mtopir}) combined with $m_t=(175\pm6)$ GeV gives us a lower
bound on $\tan\beta$: $\tan\beta\simgt1.4$
\footnote{This bound can be slightly lowered to $\tan\beta\simgt1.1$,
in gauge mediated SUSY breaking models, due to the presence of additional 
coloured matter fields at the intermediate scale $M\sim10^5-10^7$ GeV.}  
Similar results are also found in the large $\tan\beta$ region
in which both the top and bottom Yukawa couplins are 
large\footnote{In this case the bottom pole mass may  be significantly 
different from  the running mass due to the supersymmetric loop corrections 
\cite{HARASA,CAOLPOWA}. Similar corrections may be even important for the 
Kobayashi-Maskawa mixing angles \cite{BLPORA}}.
In the numerical calculation two-loop RG equations are used. 

It is clear from eq. (\ref{eqn:mtopir}) that the experimental value of 
$m_t\approx 175$ GeV
is very close to its perturbative upper bound in the MSSM. To know 
how close, we need to know $\tan\beta$ but one is tempted to speculate
that $m_t\approx m^{QF}_t$ and then $\tan\beta$
must be  either close to its lower bound or very large. 

An interesting observation is that the GUT assumption about unification
of the bottom and tau Yukawa couplings gives independent support to the
idea that $Y_t(M_Z)=Y^{QF}_t(M_Z)$ \cite{AR,CAOLPOWA1}. 
Quantitatively, this conclusion
depends on the values of $\alpha_3(M_Z)$ and $m_b(pole)$ and on the
threshold corrections to the relation $Y_b=Y_{\tau}$. Generically, however,
strong interaction renormalization effects for $Y_b$ are too strong and  
large top quark Yukawa coupling contribution to the running of $Y_b$ is 
needed to balance them (see Fig. \ref{fig:irfix}b). 
For $Y_b=Y_{\tau}$ within 10\%, $\alpha_3(M_Z)>0.11$ 
and $m_b(pole)<$5.2 GeV one gets $m_t\approx m_t^{QF}$ within 
10\%\footnote{For large $\tan\beta$ this conclusion may be less strong.}. 
Thus, the possibility of $m_t\approx m_t^{QF}$
is supported by several independent arguments (also models for dynamical
determination of $Y_t$ give values close to the IR fixed point \cite{KUZWPA})
and, together with the measured value of $m_t=175\pm6$ GeV, it makes the 
region $1.5\simlt\tan\beta\simlt2$ particularly interesting.

Another interesting region of $\tan\beta$ is $\tan\beta\approx m_t/m_b$.
Large values of $\tan\beta$ have been discussed for some time as a solution 
to the $m_t/m_b$ hierarchy with full unification of the third generation 
Yukawa couplings $Y_t=Y_b=Y_\tau$ \cite{OLPO,BA}. Such unification is 
predicted e.g. by simple versions of the SUSY $SO(10)$ GUT 
\cite{OLPO,ANRADIHA} and remains to be an interesting possibility.

In conclusion, although one cannot rule out intermediate values of $\tan\beta$,
the arguments given above are compelling enough to consider the low and large 
$\tan\beta$ regions as ``interesting'' ones. In those regions the  Higgs
boson masses are bounded more strongly than in the general MSSM. 
For $m_t=175$ GeV, from Fig. \ref{fig:mh}b we see that if $Y_t$ is 
$\simlt10$\% away from its IR value (which corresponds
to $\tan\beta\simlt1.85$) we have $M_{h^0}\simlt100$ GeV.
For large $\tan\beta$ we have $M_{h^0}\approx M_{A^0}$ up to 
$\sim 120$ GeV and for heavier $M_{A^0}$ (for arguments in favour of such 
$M_{A^0}$, see the previous Section) the $M_{h^0}$ remains constant
(independent of $M_{A^0}$) with the value fixed by radiative corrections. 

Finally, we are going to discuss several constraints on the range
of the soft supersymmetry breaking masses $m^2_1$, $m^2_2$, $m^2_3$, the 
top squark masses and the mixing parameters $A_t$, $\mu$ (i.e. the remaining 
parameters relevant for $M_h$) which follow from the extrapolation of the
MSSM to high energies. They are particularly interesting and easy to discuss
under the assumption that the $h_t$ is not too far from its quasi-infrared 
fixed point limit. 
For the sake of definitness we shall mainly focus on the low and 
intermediate $\tan\beta$ region and present analytic results in the 1-loop 
approximation \cite{CAOLPOWA1,CACHOLPOWA}. More complete 2-loop numerical 
calculations \cite{CAOLPOWA1,CAWA} confirm very well these analytic considerations. 

In the 1-loop approximation and expanding in $y(t)\equiv Y_t(t)/Y^{QF}_t(t)$ 
the RG equations for the dimensionful parameters can be solved analytically 
\cite{CAOLPOWA1,CACHOLPOWA}. Denoting by
$\ovm^2(t)\equiv m^2_Q(t) + m^2_U(t) + m^2_{H_2}(t)$,
$2\pi t\equiv\log{M\over Q}$ with $M=M_{GUT}$ or 
any intermediate scale, and with $m_K^2$, $K = H_i,Q,U,D,L,E$, standing for
the soft supersymmetry breaking mass parameters of the Higgs, left-handed 
squark, right-handed up-type squark, right-handed down-type squark, 
left-handed slepton and right-handed slepton, respectively we get the 
the following results:
\begin{eqnarray}
\ovm^2(t) &=& (1-y)\ovm^2(0) - y(1-y)A_t(0)\left(A_t(0)-2\hat\xi M_{1/2}\right)
\nonumber\\
&+& (\overline\eta - y\hat\eta + y^2\hat\xi^2) M_{1/2}^2,
\label{eqn:m2ov0}
\end{eqnarray}
\begin{eqnarray}
m_K^2(t) &=&m_K^2(0) - {c_K\over c_t}y\ovm^2(0)
- {c_K\over c_t} y(1-y)A_t(0)\left(A_t(0)-2\hat\xi M_{1/2}\right)
\nonumber\\
&+&\left[\eta_K-{c_K\over c_t}\left(y\hat\eta+y^2\hat\xi^2\right)
\right]M_{1/2}^2 + D_K,
\label{eqn:m2k0}
\end{eqnarray}
\begin{eqnarray}
D_K = -\kappa_K \ovm^2_Y(0)
\left[1-\left({\alpha_1(0)\over\alpha_1(t)}\right)^{-13/33}\right],
\label{eqn:dterm}
\end{eqnarray}
\begin{eqnarray}
\ovm^2_Y(t)&\equiv& -m^2_{H_1}(t) + m^2_{H_2}(t) \nonumber\\
&+&\sum_{gen}\left[m^2_E(t)- m^2_L(t) 
+ m^2_Q(t) + m^2_D(t) -2 m^2_U(t)\right]. 
\label{eqn:m2yovt}
\end{eqnarray}
Functions $\eta_K(t)$, $\hat\xi(t)$, $\hat\eta(t)$ 
($\overline\eta\equiv\eta_Q+\eta_U+\eta_{H_2}$) are given in closed forms in
terms of integrals over the gauge couplings and are 
defined in the Appendix of ref. \cite{CACHOLPOWA}. For $M=2\times10^{16}$ GeV
and $\alpha_3(M_Z)=0.12$ they take values: $\hat\xi=2.23$, $\hat\eta=12.8$,
$\eta_Q=7.04$, $\eta_U=6.62$, $\eta_{H_1}=\eta_{H_2}=0.513$ 
The coefficients $c_K$ and $\kappa_K$ read: 
$c_Q=1$, $c_U=2$, $c_{H_2}=3$, $c_L=c_E=c_D=c_{H_1}=0$; 
$\kappa_{H_1}=-\kappa_{H_2}=\kappa_L = -3/26$, $\kappa_E = 3/13$, 
$\kappa_Q = 1/26$, $\kappa_U = -2/13$, $\kappa_D = 1/13$. 
The evolution of the trilinear couplings $A_k$ are given by:  
\begin{eqnarray}
A_i(t) = A_i(0) - {C_i\over c_t} y A_t(0) 
+ \left({C_i\over c_t}y\hat\xi - \xi_i\right) M_{1/2},
\label{eqn:at0}
\end{eqnarray}
Here $C_t=c_t=6$, $C_b=1$ and $C_\tau=0$.
Factors $\xi_i(t)$ 
are defined in the Appendix of 
ref. \cite{CACHOLPOWA} and for $M=2\times10^{16}$ GeV and $\alpha_3(M_Z)=0.12$ 
$\xi_t=3.97$.
Quantities at $t=0$ are the initial values of the parameters at the 
scale $M$, $M_{1/2}\equiv M_3(0)$ is the initial gaugino (gluino) 
mass (computing numerical values of $\hat\xi$, $\hat\eta$... we have
assumed that $M_1/\alpha_1(0)=M_2/\alpha_2(0)=M_3/\alpha_3(0)$). 

There are several interesting observations about solutions 
(\ref{eqn:m2ov0}-\ref{eqn:at0}). 
Firstly, to a very good approximation squark mass
parameters of the first two generations decouple from the running of the
masses $m^2_{H_1}$ and $m^2_{H_2}$ (they enter only through small hypecharge
$D$-term (\ref{eqn:dterm})). As stressed in refs. \cite{DIGI,CACHOLPOWA},
this is very important for the ``naturalness'' problem. Moreover, we observe 
interesting ``fixed point'' behaviour for the parameter $A_t$ 
which, in the limit $y\rightarrow1$, becomes independent of its initial 
values and fixed in terms of the gaugino mass $M_{1/2}$ \cite{CAOLPOWA1}
(unless $A_t(0)\gg M_{1/2}$ but large values of $A_t(0)$ are
constrained by the requirement of the absence of the colour breaking minima).

Having relations (\ref{eqn:m2ov0}-\ref{eqn:at0}) we can discuss the impact 
of the requirement of the proper electroweak symmetry breaking on the low
energy parameter space, under various assumptions about the pattern of the
soft supersymmetry breaking parameters at large scale. This requirement 
correlates the low energy soft supersymmetry breaking masses in the Higgs
potential with the values of $M_{Z^0}$ and $\tan\beta$, as given in eqs.
(\ref{eqn:vsquared},\ref{eqn:sinbeta}) and, in turn, with other parameters 
which enter into their RGE.
Actually, under each specific assumption about the scale of supersymmetry
breaking and the pattern of the soft terms, one can perform a global
analysis which includes the electroweak breaking and the existing experimental 
constraints, and obtain the predictions for the Higgs sector. We present
here the results of such an analysis for the following three scenarios:
two supergravity scenarios with the GUT relation for the gaugino masses 
$M_1=M_2=M_3\equiv M_{1/2}$: one with universal soft scalar masses (and 
universal $A$-terms) and one with universal scalar masses in $SO(10)$ 
multiplets i.e. with universal sfermion masses (and universal $A$-terms)
and two soft Higgs boson masses as independent parameters,
and the gauge mediated supersymmetry breaking scenario. Part of the most 
important experimental constraints has already be discussed in Section 3.
In the global analysis discussed now, we also include the following bounds:
$m_{C^\pm}>85$ GeV, $\Gamma(Z^0\rightarrow N^0_1N^0_1)<4$ MeV,
$BR(Z^0\rightarrow N^0_1N^0_2)<10^{-4}$. 

There has been often addressed the question of fine-tuning (large 
cancellations) in the Higgs potential in models with the soft terms generated
at large scales \cite{BAGI,DIGI,ANCA}. 
Indeed, if supersymmetry is to be the solution to the 
hierarchy problem in the SM, it should not introduce another fine-tuning
in the Higgs potential. The origin of the problem is easy to see. 
Combining (\ref{eqn:vsquared}) and (\ref{eqn:m2ov0}-\ref{eqn:at0}) 
we can express $M_{Z^0}$ for a given $\tan\beta$ 
in terms of the initial values $m^2_K(0)$, $M_{1/2}$ and the $\mu$ 
parameter:
\begin{eqnarray}
M^2_{Z^0}&=&-2\mu^2(t) + a_{H_1}m^2_{H_1}(0) + a_{H_2}m^2_{H_2}(0)
+ a_{QU}\left(m^2_Q(0)+m^2_U(0)\right)\nonumber\\
&+& a_{AA}A_t^2(0) + a_{AM}A_t(0)M_{1/2}
+ a_MM^2_{1/2}
\label{eqn:zmass}
\end{eqnarray}
For $m_t=175$ GeV the generic values of the coefficients in 
eq. (\ref{eqn:zmass}) in the supergravity scenario e.g. 
for $\tan\beta\approx1.65(2.2)$ corresponding to $y\approx0.95(0.85)$
are $a_{H_1}\approx1.2(0.5)$,
$a_{H_2}\approx1.7(1.5)$, $a_{QU}\approx1.5(1.1)$, $a_{AA}\approx0.1(0.2)$, 
$a_{AM}\approx-0.3(-0.7)$, $a_M\approx15.0(10.8)$.
Eq. (\ref{eqn:zmass}) shows that for values of $\mu$, $M_{1/2}$ and/or 
$m^2_K(0)$ much larger than $M_{Z^0}$ one needs large 
cancellations. Asymptotically, we are back to the hierarchy problem in the SM.
Although the idea of ``naturalness'' is only qualitative, one can at least 
correlate the magnitude of the necessary cancellations with the values of
the parameters $\mu$, $m^2_K(0)$ and $M_{1/2}$ and, in consequence,
with the low energy mass parameters. One notes, in particular, that 
the smalness of $a_{QU}$ (compared to $a_M$) puts weaker constraints on the
``natural'' values of $m^2_K(0)$ than large $a_M$ does on $M_{1/2}$. 
However, in the physical spectrum this effect is partially counterbalanced
by the fact that the stop soft masses tend to be suppressed compared to 
$m_{Q,U}(0)$ by the running with large top quark Yukawa coupling.
This effect is stronger for the right handed stop than for the left handed 
one and gives the hierarchy $M_{\tilde t_R}<M_{\tilde t_L}$. 
Important source of fine-tuning can also be the relation (\ref{eqn:sinbeta})
which correlates the values of $\tan\beta$ and the $B_0$ parameter.

``Naturalness'' of a given parameter set can be quantified e.g. by calculating 
the derivatives of $M^2_{Z^0}$ with respect to the soft mass parameters
\cite{BAGI,DIGI}:
\begin{eqnarray}
\Delta_i\equiv\mid{a_i\over M_Z^2}{\partial M^2_Z\over\partial a_i}\mid
\label{eqn:der}
\end{eqnarray}
(other criteria have also been proposed \cite{ANCA} but will not be 
discussed here since, at any rate, the concept of naturalness is only
qualitative.) 

In the global analysis, which combines the electroweak breaking with 
experimental constraints, it is interesting to check the ``naturalness'' 
of different parameter regions i.e. to check the values $(\Delta_i)$
for each parameter set. Before presenting the results for the three scenarios 
considered, it is worthwile to remember several general remarks. First, as
already said, the ``naturalness'' criterion is only a qualitative one and
it is unclear how big cancellations are ``acceptable'' ones. The
hierarchy problem in the SM means fine-tuning of many orders of magnitude
and from that perspective cancellations of one, two or even three orders
of magnitude are still very small. Secondly, we do not know the theory in
which soft supersymmetry breaking terms originate and it may well be that 
such a theory will give them correlated to each other, thus ``explaining''
the cancellations between them. Finally one can see from eqs. 
(\ref{eqn:vsquared},\ref{eqn:sinbeta}) and (\ref{eqn:zmass}) that the
necessary cancellations tend to increase in the small and large $\tan\beta$
regions advocated as the most interesting ones on the basis of earlier 
arguments. Indeed, the coefficients $a_i$ in eq. (\ref{eqn:zmass}) have
$1/(\tan^2\beta-1)$ singularity. Moreover, both for small and very
large $\tan\beta$ eq. (\ref{eqn:sinbeta}) gives very strong dependence
on $B_0$ (i.e. large derivative with respect to $B_0$ in eq. (\ref{eqn:der})).

With those comments in mind we now present the results of our
global analysis, for the three scenarios considered and for several values
of $\tan\beta$. In each case the lightest Higgs boson mass is shown as a
function of the heavier stop and the maximal and minimal values of
$(\Delta_i)^{max}$ for the points on each plot are also noted.

\begin{figure}
\psfig{figure=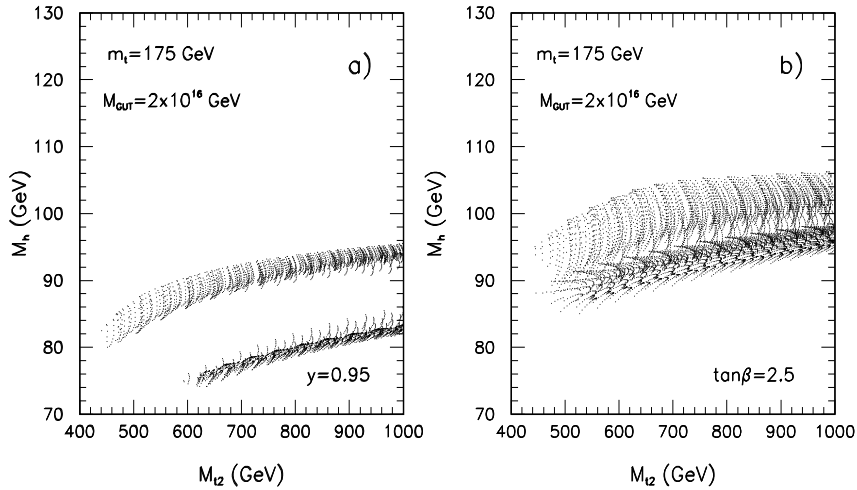,width=12.0cm,height=7.0cm}
\caption{$M_{h^0}$ versus the mass of the heavier stop ($M_{\tilde t_2}$)
obtained from universal boundary conditions at the scale $2\times10^{16}$ GeV
{\bf a)} close to the Q-IR fix point of the top quark mass ($y\approx0.95$) ~
${\bf b)}$ for $\tan\beta=2.5$ ($y\approx0.8$). In both cases
no fine-tunig criterion is imposed.
\label{fig:fintu1}}
\end{figure}

Widely discussed has been
the so-called minimal supergravity model ({\sl An\-satz}) 
with universal scalar and gaugino masses and universal 
trilinear soft terms.  In this model all superpartner masses are given in 
terms of five parameters: $m_0^2$, $M_{1/2}$, $\mu$, $A_0$ and $B_0$. Two of
them can be traded for $M_{Z^0}$ and $\tan\beta$. Thus, we get
strongly correlated superpartner spectrum and correlated with the Higgs boson
masses. It is now particularly simple to follow our global analysis and to
determine the allowed range of the lightest
Higgs boson mass as a function of the heavier stop mass.
In Fig. \ref{fig:fintu1} we show the results for $\tan\beta=1.65$ and 2.5
(corresponding to $y\approx0.95(0.80)$).
The parameter space has been scanned up to $M_{\tilde t_2}=1$ TeV with
$\mu_0$ and $B_0$ fixed by $M_{Z^0}$ and $\tan\beta$. We see that, in this 
model, requiring the proper breaking of the electroweak symmetry
and with the imposed experimental constraints the lightest Higgs 
boson mass is bounded from below: $M_{h^0}\simgt75(85)$ GeV for 
$\tan\beta=1.65(2.5)$ (for $\tan\beta=10$ the lower bound is around 105 GeV). 

The model also gives lower bound on $M_{A^0}$ of about 500 GeV at 
$\tan\beta=1.65$ and decreasing to 300 GeV at $\tan\beta=10$. The heavier stop 
is bounded from below at $\sim450$ GeV. Of course, the crucial role in 
obtaining those bounds is played by the universality {\sl Ansatz} combined 
with the existing experimental constraints. We note also an interesting 
dependence on the sign of the $\mu$ parameter (two clear branches in Fig. 
\ref{fig:fintu1}a) which is a reflection of the acceptable region in the 
$(\theta_{\tilde t}, M_{\tilde t_2})$ plane shown in Fig. \ref{fig:bsgamma}b.
The mass $M_{h^0}$ is bounded from above at 95, 105 and 120  GeV for
$\tan\beta=1.65$, 2.5 and 10, respectively. Thus the general bounds shown in
Fig \ref{fig:mh}b can be reached even in this constrained model.

\begin{figure}
\psfig{figure=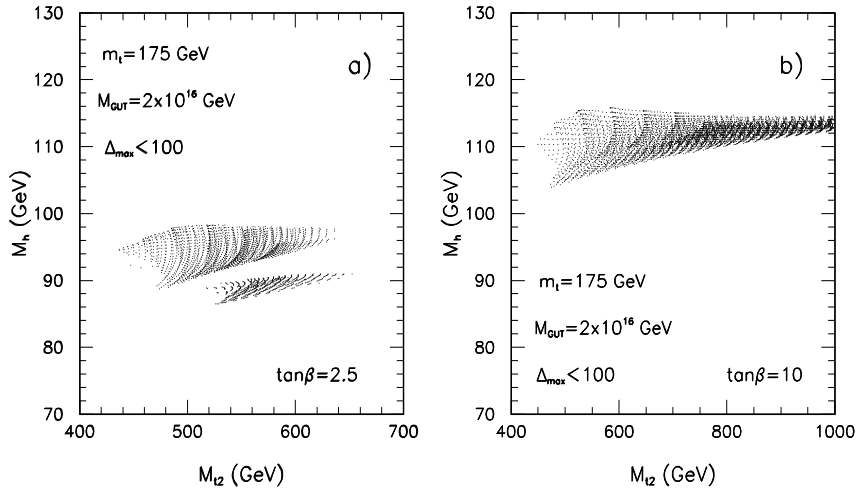,width=12.0cm,height=7.0cm}
\caption{$M_{h^0}$ versus the mass of the heavier stop ($M_{\tilde t_2}$)
obtained from universal boundary conditions at the scale $2\times10^{16}$ GeV
and requiring $\Delta_i<100$ 
{\bf a)} for $\tan\beta=2.5$ ($y\approx0.80$) ~
${\bf b)}$ for $\tan\beta=10$ ($y\approx0.71$).
\label{fig:fintu2}}
\end{figure}

Turning now to the fine-tuning problem we observe first that the model does
not admit at all solutions with all $\Delta_i<10$. This is mainly because 
of the imposed experimental limit $m_{C^\pm}>85$ GeV \cite{DIGI} which
pushes $M_{1/2}$ into the region with $\Delta_{M_{1/2}}\simgt10$ for all 
$\tan\beta$ values 
\footnote{Strictly speaking, this conclusion is valid as long as we work with 
the tree level Higgs potential renormalized at $M_{Z^0}$. It is well known 
\cite{OLPO1,CACA,ANCA,DIGI} that inclusion of the full 1-loop corrections 
to the scalar potential diminishes somewhat the degree of fine-tuning and 
the bound $m_{C^\pm}>85$ GeV becomes marginally consistent with $\Delta_i<10$,
for intermediate values of $\tan\beta$. We neglect 
this effect here, as it does not change our qualitative conclusions.}.
Moreover, close to the IR fixed point (for 
$\tan\beta\approx1.65$), there do not even exist solutions with all
$\Delta_i<100$ \cite{CAOLPOWA1}. Actually, defining two different $\Delta$'s:
$\Delta_{max}^\prime\equiv max\{\Delta_{M_{1/2}}, ~\Delta_{m_0}, ~
\Delta_{\mu_0}\}$ and $\Delta_{max}\equiv max\{\Delta_{max}^\prime, ~
\Delta_{B_0}, ~\Delta_{A_0}\}$, the points in Fig. \ref{fig:fintu1}a(b) 
give $70(28)\simlt\Delta_{max}^\prime\simlt970(560)$, ~
$130(35)\simlt\Delta_{max}\simlt5400(750)$. As expected
from the general arguments, cancellations become weaker with increasing
$\tan\beta$. In Fig. \ref{fig:fintu2}a (b) we show the results for 
$\tan\beta=2.5(10)$ with the cut $\Delta_{max}<100$. We note that in this case
such a cut leaves a non-empty parameter region but gives stronger upper 
bounds on the Higgs boson mass for the same values of $M_{\tilde t_2}$. 
They result mainly from the bound on $A_0$ (i.e. on the left-right mixing)
obtained due to increasing $\Delta_{M_{1/2}}$ with increasing $A_0$.
Moreover, the cut $\Delta_{max}<100$ gives also an upper bound on 
$M_{\tilde t_2}$. A weaker cut, $\Delta_{max}^\prime<100$, does not change 
the results for $\tan\beta=10$ (as expected) but allows for broader
range of $M_{\tilde t_2}$ for $\tan\beta=2.5$ (with the upper bound for
$M_{h^0}$ as for $\Delta_{max}<100$). Finally we note one more interesting
effect: a cut on $\Delta$'s gives almost flat (instead of logarithmic)
dependence of $M_{h^0}$ on $M_{\tilde t_2}$. An increase in
$M_{\tilde t_2}$ is balanced by a decrease in $A_0$ (i.e. in the
left-right mixing) to keep $\Delta$'s below the imposed bound.

In the next step one can study a less restrictive model, with the pattern
of the soft terms consistent with the $SO(10)$ unification, i.e. with the
universal sfermion masses and the two Higgs boson masses taken as 
independent parameters. It turns out that the predictions the $M_{h^0}$,
$M_{A^0}$ and $M_{\tilde t_2}$ (as well as the degree of fine-tuning)
are very similar to those in the universal case and need not be independently
shown here. This similarity can be partly understood in terms of the 
important role played by the limit $m_{C^\pm}>85$ GeV and by the 
constraints from $b\rightarrow s\gamma$ and from precision data, which
are not sensitive to the assumed non-universality in the Higgs boson mass
parameters. Moreover, there is no real change in the values of $\Delta_i$'s
since their expected decrease with the increasing number of free 
parameters is now reduced by the dissapearance of certain cancellations
in $\Delta_i$'s which are present in the universal case. Finally, the 
considered values of $\tan\beta$ are enough above 1 to make the 
electroweak symmetry breaking easy even in the universal case and the
considered breakdown of full universality does not significantly enlarge
the parameter space consistent with the electroweak symmetry breaking. Thus, 
the results presented in Figs. \ref{fig:fintu1} and \ref{fig:fintu2} are 
representative also for the considered partial breakdown of universality.

The difference between universal and non-universal
boundary conditions becomes more visible in the large $\tan\beta$ region
\cite{OLPO2,HARASA}. In the latter case the parameters $\Delta$ are
typically $\sim\tan\beta$ whereas in the former - ${\cal O}(1000)$, i.e.
fine-tuning of order $10^{-3}$. From the point of view of the predictions for
the MSSM spectrum, the large $\tan\beta$ region is characterized by the
expectation of a relatively light $CP$-odd Higgs boson, 
$M_{A^0}\simlt {\cal O}(200$ GeV). For further details in the large $\tan\beta$
region, in particular for correlations with radiative corrections to the
$b$-quark mass and with $b\rightarrow s\gamma$ decay, we refer the reader
to \cite{OLPO2,HARASA,CAOLPOWA}. 

Finally, it is interesting to compare the supergravity scenario with models 
in which supersymmetry breaking is transmitted to the
observable sector through ordinary $SU(3)\times SU(2)\times U(1)$ 
gauge interactions of the so-called messenger fields at scales $M\ll M_{GUT}$
\cite{DIFISR}.
In general, these gauge-mediated models of SUSY breaking are characterized
by two scales: the scale $M$, which is of the order of the average messenger
mass and the scale $\sqrt{F}$ ($\sqrt{F}<M$) of supersymmetry breaking.
Messenger fields are assumed to form complete {\bf 5}+$\overline{\rm\bf 5}$ 
(or {\bf 10}+$\overline{{\rm\bf 10}}$)  $SU(5)$ representations. 
Their number $n$ is restricted to $n_{max}=4$ by the requirement of 
perturbativity of the gauge couplings up to the GUT scale.
In those models the LSP is 
a very light gravitino ($m_{\tilde G}<1$ keV). For $\sqrt{F}<10^6$ GeV the 
decay length of the lightest neutralino into a photon and gravitino is such
that this decay occurs within a typical detector. Hence, photons + missing 
energy become a signature of supersymmetry at LEP and Tevatron colliders.
The absence of such events in the existing data strongly disfavours charginos
and stops with masses below $\sim125$ GeV and $\sim140$ GeV 
respectively \cite{AMKAKRMAMR}. 

In terms of $M$ and $x\equiv F/M^2$ the soft supersymmetry breaking parameters 
of the MSSM at the scale $\sim M$ are given by:
\begin{eqnarray}
M_i = {\alpha_i(M)\over4\pi}M ~n ~x ~g(x) \equiv {\alpha_i(M)\over4\pi}M ~y
\end{eqnarray}
\begin{eqnarray}
m^2_{\tilde f} = 2M^2~n ~x^2 ~f(x) 
\sum_i \left({\alpha_i(M)\over4\pi}\right) C_i
=2M^2 ~y^2 ~z \sum_i \left({\alpha_i(M)\over4\pi}\right) C_i
\label{eqn:scalars}
\end{eqnarray}

\noindent
where $C_3=4/3$, $C_2=3/4$, $C_1=(3/5)Y^2$ ($Y$ being the hypercharge
of the scalar ${\tilde f}$), the functions $g(x)$ and
$f(x)$ ($g(0)=f(0)=1$, $g(1)\approx 1.4$, $f(1)\approx0.70$) 
can be found in ref. \cite{DIGIPO} and the factor $z\equiv f(x)/n g^2(x)$. 
Thus, for fixed messenger sector
(i.e. fixed $n$) and fixed scale $M$ all soft supersymmetry breaking masses
are predicted in terms of $y$ ($0<y<n_{max}~g(1)\approx5.6$) 
\footnote{Nonzero lower bound on $y$ 
is set by Tevatron limit on gluino mass, $m_{\tilde g}\simgt150$ GeV,
which is, however, subject to some restrictions}.
In those models we also have $A_0\approx0$ as the $A_0$ parameter can be
generated at two loop only \cite{DITHWE}. 
However, the values of the soft masses $m^2_{H_{1,2}}$ may differ 
significantly from their values given by eq. (\ref{eqn:scalars}) since they
can be modified by physics involved in generation of $B_0$ and $\mu_0$
parameters \cite{DVGIPO}. Therefore, in our scans we take $y$, $m_{H_1}$,
$m_{H_2}$, $\mu_0$ and $B_0$ as free parameters (the last two are fixed
by $M^2_{Z^0}$ and $\tan\beta$). To be general, the factor $z$ in eq. 
(\ref{eqn:scalars}) is scanned between $z_{min}=f(1)/n_{max}g^2(1)\approx0.15$ 
and $z_{max}=1$. For definitness we will consider $M=10^5$ GeV only.

Here we follow the same simple approach we used for the supergravity models. 
On the parameter space consistent with the electroweak symmetry breaking we
impose the discussed earlier experimental constraints
(now we require $m_{C^\pm}>120$ GeV, $M_{\tilde t_1}>140$ GeV). 
Very important r\^ole is played by $b\rightarrow s\gamma$. The requirement 
of good $b\rightarrow s\gamma$ rate reduces otherwise rather 
widely spread out $h^0$ and $A^0$ Higgs boson masses (for $\tan\beta=2.5$: 
$20<M_{h^0}<100$ GeV) to a narrow band ($80<M_{h^0}<100$ GeV, $M_{A^0}>200$ 
GeV). This effect can be easily understood (see Fig. \ref{fig:bsgamma}) 
because in the model considered squarks and charginos are rather heavy 
\footnote{In addition, because $\mu$ values required by electroweak
symmetry breaking are large, the lighter chargino turns out to have only
small higgsino component and hence its $b \tilde t C^-$ coupling
is weaker than that of the pure higgsino chargino which is responsible for  
the limit shown in Fig. \ref{fig:bsgamma}.} so a light $A^0$ is not allowed
by $b\rightarrow s\gamma$
and light $h^0$ is always associated with light $A^0$. Moreover, surviving 
small values of $M_{h^0}$ ($\sim80$ GeV for $\tan\beta=2.5$ are associated 
with lowest values of $M_{\tilde t_2}$ ($\simlt 500$ GeV) which are eliminated 
by imposing the $\Delta\chi^2<4$ cut. 
Finally, if we also require ``naturalness'' e.g. by demanding 
\footnote{Naturalness of the gauge mediated models has been analyzed by different
methods in refs. \cite{CIST}.} $\Delta_{max}<100$ 
($\Delta_{max}=max\{\Delta_x, ~\Delta_{m_{H1}}, ~\Delta_{m_{H2}}, ~
\Delta_{\mu_0}, ~\Delta_{B_0}\}$), we constrain the heavier stop mass 
$M_{\tilde t_2}$ and $CP$-odd higgs boson mass $M_{A^0}$ from above to $\simlt700$ 
GeV. 

\begin{figure}
\psfig{figure=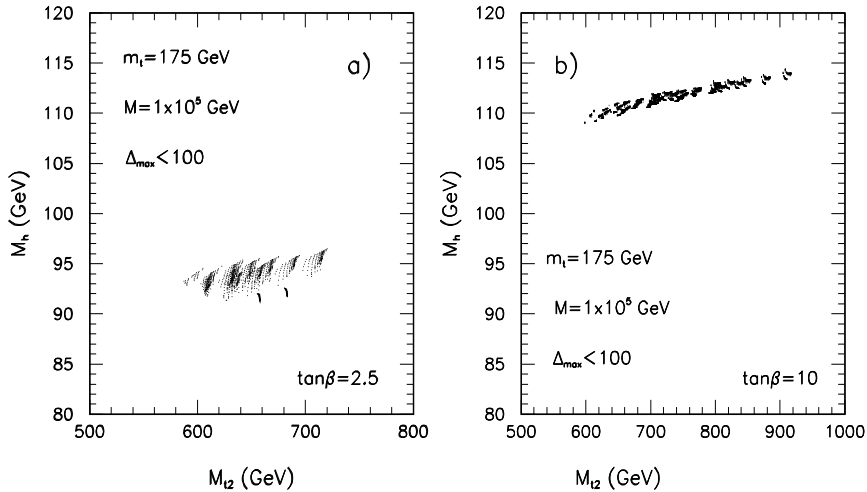,width=12.0cm,height=7.0cm}
\caption{Results for gauge mediated supersymmetry breaking models with 
$M=10^5$ GeV for $\tan\beta=2.5$: ~a) 
$M_{h^0}$ versus the mass of the heavier stop ($M_{\tilde t_2}$)
${\bf b)}$ versus the mass of the $CP$-odd Higgs boson ($M_{A^0}$).
Condition $\Delta_i<100$ is imposed. 
\label{fig:fintu3}}
\end{figure}

Final results are shown in 
Fig. \ref{fig:fintu3} as a plot of $M_{h^0}$ 
versus the mass of the heavier stop $M_{\tilde t_2}$ predicted in models 
of gauge mediated supersymmetry breaking with $M=10^5$ GeV for 
$\tan\beta=2.5$ and 10. 
As in the case of supergravity models, the restriction of the chargino and 
stop masses eliminates solutions with $\Delta_{max}<10$. 
With all constraints imposed, 
$M_{h^0}$ turns out to be surprisingly tightly constrained. 
For $\tan\beta=2.5(10)$ values of the lightest scalar Higgs boson
are bounded by 90(108) GeV $\simlt M_{h^0}\simlt97(115)$ GeV.
Masses of the $CP$-odd Higgs boson are bounded to 280(200) GeV
$\simlt M_{A^0}\simlt$700(850) GeV. These upper bounds should be compared 
to the ones obtained in \cite{RITOMO} in the restricted model of
gauge mediated supersymmetry breaking (with $x=1$, $n=1$ and with
$m_{H_{1,2}}$ as given by eq. (\ref{eqn:scalars})) without imposing any
additional constraints. It is interesting that in the much more general 
scenario described above, after imposing experimental and 
naturalness cuts, 
one gets upper bounds on $M_{h^0}$ not higher than those obtained 
in \cite{RITOMO}.

\section{Non-minimal SUSY models}
Non-minimal supersymmetric extensions of the Standard Model can go beyond
the MSSM in several directions. Models with additional Higgs singlets
\cite{FA,DR}, doublets and/or triplets have been considered in the literature 
\cite{ESQU1}.
The gauge group of the MSSM can also be extended by e.g. additional $U(1)$ 
factors. Finally, models with $R-$parity spontaneously broken (Supersymmetric
Singlet Majoron Model) have been proposed \cite{GIMAPIRI}. The important 
question is what can be said about the Higgs sector in such extensions. 

In the most popular extension, the so called Next-to-Minimal Supersymmetric
Standard Model (NMSSM) \cite{FA,DR}, one 
introduces a gauge singlet superfield $N$ and replaces the term 
$\mu H_1 H_2$ in the MSSM superpotential by
\begin{eqnarray}
\lambda H_1 H_2 N + {\kappa\over3} N^3
\end{eqnarray}
with the obvious motivation to avoid the $\mu$ problem \cite{HALY}.
The model has been analysed in several papers.
At the tree level, the upper bound for the mass of the lightest Higgs boson 
\begin{eqnarray}
M^2_{h^0}\leq M_Z^2\left(\cos^22\beta + {2\lambda^2\over g^2} c^2_W
\sin^22\beta\right)
\end{eqnarray}
has been derived \cite{DR,ESQU1}. 
It is important that this bound  depends neither
on the sfermion masses (which can be much larger than $M_Z$) nor on the
vacuum expectation value of the singlet field $N$ (which is not constrained
by $M_Z$). Therefore, the 
bound is controlled only by the dimensionless Yukawa coupling $\lambda$ which 
is constrained by the requirement of the perturbativity of the model, 
$\lambda\simlt4\pi$. 

This result, that the tree level bound does not depend on the parameters
which can be of order of the supersymmetry breaking scale (and, hence, is 
always $\simlt {\cal O}(G_F^{-1/2})$), has been extended to 
models containing arbitrary numbers of Higgs 
singlets, doublets and triplets \cite{ESQU1,ESQU2}. Finally it was also
shown to be valid in the Supersymmetric Single Majoron Model \cite{ES}. 
Almost model-independent 
proofs of this remarkable fact has been given in \cite{KAKOWE,ES}.
Radiative corrections to the tree level bound \cite{ELRA}
are under control as in the MSSM and has been
shown to depend on soft SUSY breaking masses only logarithmically
\cite{ELKIWH1,ELKIWH2,ES}. 
 
The tree level bounds can be further strenghtened by
requiring that the additional Yukawa couplings remain perturbative up to
a scale $\Lambda$ (GUT scale). For the NMSSM this has been done in the papers
\cite{ESQU1,BISA,ELKIWH2}. Analysing 
coupled RGE for $\lambda$, $Y_t$ and $\alpha_s$
the upper limit on $\lambda$ has been found as a function of $Y_t$ for 
$\Lambda=10^{16}$ GeV. For large values of $m_t$, for which $Y_t$ is always
close to its perturbativity limit, the coupling $\lambda$ is forced to be 
small and hence its effects become less important.
Including all relevant one-loop radiative corrections 
it was found that, for $m_t$ in the present experimental range,  the upper
bound for the lightest Higgs boson in the NMSSM is only $5-15$ GeV larger
than the corresponding upper bound in the MSSM and, for $M_{\tilde t_i}\leq1$
TeV it is smaller than 150 GeV.
Effects of the additional assumptions like gauge coupling unification,
universality of the soft SUSY breaking masses and requirement of the correct
electroweak symmetry breaking have also been studied in this model \cite{KIWH}.

\section{Summary}
In unconstrained minimal supersymmetric extension of the Standard Model
there exists the well known upper bound on the mass of the lightest 
supersymmetric Higgs. The available parameter space of the model is now 
considerably reduced by the existing experimental data and can be further
reduced by additional theoretical assumptions, mostly related to the 
extrapolation of the model to very large energy scales. Especially fruitful
is the assumption about perturbative validity of the model up to the GUT 
scale and the requirement of the proper electroweak breaking combined
with a simple {\sl Ansatz} (such as universality or partial universality)
for the pattern of the soft supersymmetry breaking terms.
Such a reduction in the parameter space results in more definite
expectations for $M_{h^0}$ than the general bounds. Several
arguments point toward $M_{h^0}<$100 GeV. Both, the discovery or the
absence of the Higgs boson in this mass range will have strong implications
for the supersymmetric extension of the Standard Model.

\section*{Acknowledgments}
Work supported by the Polish Commitee for Scientific Research
under the grant  2 P03B 040 12
and by the European Union under contract  CIPD-CT94-0016.
P.H.Ch. would like to thank Dr. M. Misiak for cross checking 
his code for $b\rightarrow s\gamma$.

\end{document}